\documentclass[submission,copyright,creativecommons]{eptcs}
\providecommand{\event}{PDMC 2011} 
\usepackage{breakurl}             

\usepackage{multirow}
\usepackage{xspace}

\usepackage{tikz}
\usetikzlibrary{arrows,automata,calc,trees}

\usepackage{amsmath,amsthm,amssymb}
\theoremstyle{plain}

\theoremstyle{definition}

\title{CoInDiVinE: Parallel Distributed Model Checker \\
  	for Component-Based Systems}
\author{Nikola Bene\v s\thanks{The author has been supported by Czech Grant Agency, grant no.~GD102/09/H042} 
	\qquad Ivana \v Cern\'a\thanks{The author has been supported by Czech Grant Agency, grant no.~GA201/09/1389}
	\qquad Milan K\v riv\'anek\thanks{The author has been supported by Czech Grant Agency, grant no.~GAP202/11/0312}
\institute{Faculty of Informatics\\
Masaryk University\\
Brno, Czech Republic}
}
\def\titlerunning{CoInDiVinE}
\def\authorrunning{N. Bene\v s, I. \v Cern\'a \& M. K\v riv\'anek}
\begin{document}
\maketitle

\begin{abstract}
CoInDiVinE is a~tool for parallel distributed
model checking of interactions among components 
in hierarchical compo\-nent-based systems. 
The tool extends the DiVinE framework with a~new input language 
(component-interaction automata) and a~property specification logic (CI-LTL).
As the language differs from the input language of DiVinE, 
our tool employs a~new state space generation algorithm that also
supports partial order reduction.
Experiments indicate that the tool has good scaling properties 
when run in parallel setting.
\end{abstract}

\section{Introduction}
Component-based systems engineering is nowadays an established 
software development technique. 
This practice aims at building complex software systems using autonomous
prefabricated components and assembling them in a~possibly hierarchical manner.
This approach allows for component reuse and may bring benefits in
reduction of development cost.
However, as with any other software system, correctness is often 
a~critical issue. 
Moreover, in component-based systems the correctness has two aspects:
the correct behaviour of the components themselves and the correct interaction
among them.  

To model and reason about component interactions, an~automata-based formalism
called \emph{component-interaction automata} (CI automata in the following)
has been contrived in~\cite{CIA:SAVCBS-05-SEN}.
The formalism is very general and allows to model various types of
component systems. 
The main advantage of the formalism is its generalized notion of
composition which may be used to model various kinds of component linking,
allowing even for hierarchical description of the component architecture.
The formalism is equipped with a~logic called 
CI-LTL~\cite{CoCoME} which is an extension of the classical linear temporal
logic that allows to specify properties of component interaction.
As the formalism is automata-based and the logic is based on LTL, 
this makes them naturally well suited for the automata-based 
model checking approach.

The tool we are presenting, CoInDiVinE,
verifies systems described as hierarchical compositions of 
CI automata against CI-LTL properties.
The tool extends the DiVinE tool~\cite{divine-homepage,BBCR10}
with new input languages, specifying the CI automata formalism
and the CI-LTL logic. As the modelling formalism has a hierarchical
structure, the tool is equipped with a tailored mechanism for state space
generation.  

The model checking task suffers from the state-space explosion problem
even more in the field of component-based systems. The systems
are assembled from a~large number of autonomous components,
resulting in a~high degree of concurrency and interleaving.
It is thus expected that the state space may be exponential 
in the number of components.
Our tool employs common methods (parallelism, distributed computing,
state space reduction) to fight the state-space explosion.
There are different kinds of state space reduction, such as bisimulation
minimization~\cite{CBSE:Lumpe} or partial order reduction~\cite{CIA:IFM-09}.
Our tool provides the latter.

In our previous work~\cite{CIA:PDMC-08-ENTCS} we report on a~case study where 
a~realistic component system has been modelled 
and verified with the
help of CI automata and CI-LTL logic. These experiments were performed on 
a~then current extension of the DiVinE tool (0.7)
which did not support partial order
reduction and used a~slow state space generating algorithm (see the next
section for a~discussion of the state space generating algorithms for CI
automata).  The current tool exhibits a substantial extension of the
original implementation by a new, faster and better scaling state space
generation algorithm and support for partial order reduction.



\vspace{-0.5em}

\section{Modelling Formalism and State Space Generation}

The modelling formalism of component-interaction automata was first
described in~\cite{CIA:SAVCBS-05-SEN}. 
The formalism is a~generalization of previous automata-based 
formalisms~\cite{Automata:Interface:interface_automata,Lynch89anintroduction}.
A~CI automaton is a~finite transition system whose
transitions are labelled with triples $(m,a,n)$. Each label represents
a~component interaction with $a$ being the name of the action, $m$ the identification of
the sending component 
and $n$ the identification of the receiving
component. 
At most one of $m$, $n$ can be the special symbol $-$
with the meaning that the interaction is open, i.e.~an input action
$(-,a,n)$ or an output action $(m,a,-)$. The remaining labels represent
internal communication.  

A~set of CI automata may be composed. 
All the transitions of the original CI automata may be inherited by
the composite automaton and new transitions may be created by synchronization,
combining open labels $(m,a,-)$ and $(-,a,n)$ into $(m,a,n)$. 
Which of these labels are present in the composite automaton is governed
by a~composition parameter. The set of labels allowed by the composition
parameter is called a~\emph{set of feasible labels} for the composition.
%
%
A~model of a~component-based system is described as a~hierarchy tree. 
The leaves
of the tree are primitive CI automata, whose transitions are given explicitly.
The internal nodes represent the compositions (with the composition parameter).

The state space generation proceeds on-the-fly, by repeatedly using
an algorithm for computing successor states for a~given state.
A~state of the whole system is represented as a~tuple of states of all
primitive CI automata.
We present the ideas of 
two algorithms, the (naive) recursive algorithm and a~better 
algorithm based on precomputation. 
Full description of the algorithms can be found in~\cite{milan-diplomka}.

The idea behind the recursive algorithm is straightforward.
To compute the transitions
of a~composite automaton we first obtain the transitions of its children
in the hierarchy tree. We then combine them according to the composition
parameter. The children may be either primitive automata (leaves) or
composite automata (internal nodes). For primitive automata,
the transition relation is given explicitly and obtaining outgoing
transitions of a~given state is thus simple. For composite automata,
we recursively proceed with the same algorithm.



The alternative algorithm is based on the notion of 
the \emph{lowest common ancestor}. The lowest common ancestor of two leaves $A$, $B$
in the hierarchy tree is an internal node $C$ such that the subtree rooted
at $C$ contains both $A$ and $B$ and there is no smaller subtree with 
this property.
The algorithm (LCA in the following) is based on the following
idea: 
given a~system of CI automata, 
 we need to decide whether a~transition is enabled in the current configuration or not.
 In the case of input, output or internal transition inherited from a~primitive automaton
we just follow the path between the primitive automaton and the root in the 
hierarchy tree and check if the label is allowed in all the 
composite automata on the path.

The situation is a~little more complicated when two automata synchronize.
Synchronization transitions only originate from composite automata where two
of their elementary components synchronize on complementary external
actions.  Let the labels of these external transitions be $(s, a, -)$ and
$(-, a, r)$.  First of all, we find the lowest common ancestor $\lambda$ of
the two primitive automata $S_i$ and $S_j$ in the system tree.  Next, we
have to check if the label $(s, a, -)$ from the automaton $S_i$ (w.l.o.g.)
is included in all the sets of feasible labels along the path from $S_i$ to
$\lambda$.  Similarly, the label $(-, a, r)$ from the automaton $S_j$ must
be included in every set of feasible labels along the path from $S_j$ to
$\lambda$, so that the two automata can synchronize in $\lambda$.

\begin{figure}
\footnotesize
\begin{center}
\begin{tikzpicture}[scale=0.8, transform shape]
\node[state] (ROOT) {$\phantom{C_1}$} [level distance=1.6cm,sibling distance=2.25cm]
  child [level distance=4.8cm,sibling distance=6.7cm] { node[state] {$\phantom{S_1}$} }
  child { node[state,fill=gray!20!white] (LCA) {$\phantom{C_2}$}
    child { node[state] (C3) {$\phantom{C_3}$} [sibling distance=1.5cm]
      child { node[state,fill=gray!20!white] (U) {$S_i$} }
      child { node[state] {$\phantom{S_3}$} }
    }
    child [level distance=3.2cm] { node[state] {$\phantom{S_4}$} }
    child { node[state] (C4) {$\phantom{C_4}$} [sibling distance=1.5cm]
      child { node[state,fill=gray!20!white] (V) {$S_j$} }
      child { node[state] {$\phantom{S_6}$} }
    }
  }
;

\node[right of=LCA,node distance=2cm] (PIN) {$\lambda(S_i, S_j)$};
\path (LCA) edge [<-] (PIN);

\coordinate (ROOTCS) at ($ (ROOT.center) ! 0.1cm ! (LCA.center) $);
\coordinate (ROOTP) at ($ (ROOTCS) ! 0.6cm ! 90 : (LCA.center) $);

\coordinate (LCACS1) at ($ (LCA.center) ! 0.2cm ! (ROOT.center) $);
\coordinate (LCAP1) at ($ (LCACS1) ! 0.6cm ! 270 : (ROOT.center) $);

\coordinate (LCACS2) at ($ (LCA.center) ! 0.2cm ! (C3.center) $);
\coordinate (LCAP2) at ($ (LCACS2) ! 0.6cm ! 270 : (C3.center) $);

\coordinate (C3P1) at ($ (C3) ! 0.6cm ! 90 : (LCA.center) $);
\coordinate (C3P2) at ($ (C3) ! 0.6cm ! 270 : (U.center) $);

\coordinate (UCS) at ($ (U.center) ! 0.1cm ! (C3.center) $);
\coordinate (UP) at ($ (UCS) ! 0.6cm ! 90 : (C3.center) $);

\coordinate (C3P) at (intersection of LCAP2--C3P1 and C3P2--UP);

\coordinate (LCACS3) at ($ (LCA.center) ! 0.5cm ! (C4.center) $);
\coordinate (LCAP3) at ($ (LCACS3) ! 0.45cm ! 270 : (C4.center) $);

\coordinate (C4P1) at ($ (C4) ! 0.45cm ! 90 : (LCA.center) $);
\coordinate (C4P2) at ($ (C4) ! 0.45cm ! 270 : (V.center) $);

\coordinate (VCS) at ($ (V.center) ! 0.4cm ! (C4.center) $);
\coordinate (VP) at ($ (VCS) ! 0.45cm ! 90 : (C4.center) $);

\coordinate (C4P) at (intersection of LCAP3--C4P1 and C4P2--VP);

\path (ROOTP) edge [<->,red] (LCAP1);

\path (LCAP2) edge [<-,red] (C3P)
      (C3P) edge [->,red] (UP);

\path (LCAP3) edge [<-,red] (C4P)
      (C4P) edge [->,red] (VP);
\end{tikzpicture}
\end{center}
\caption{System of CI automata}
\label{fig:hier_struct}
\end{figure}
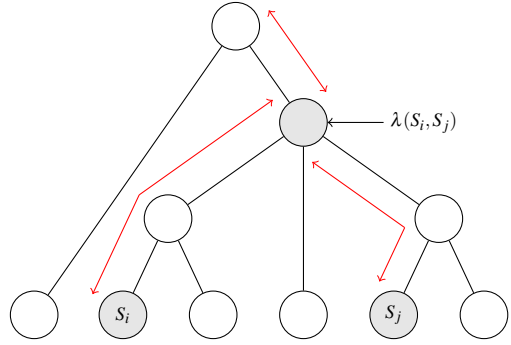

This is sufficient for a~new synchronization transition labelled $(s, a, r)$
to be formed in $\lambda$. However, this does not guarantee that it is
enabled in the resulting composition as the transition can be removed from
the transition space by subsequent compositions.  Therefore we also have to
check if the new label is included in all the sets of feasible labels from
$\lambda$ to the root of the tree. The situation is illustrated in 
Figure~\ref{fig:hier_struct}.

The state generation now has an initialization phase, which is run once
at the beginning. In the initialization phase we first compute 
the lowest common ancestors for every pair of primitive automata.
We then compute the intersection of all sets of feasible labels along 
each path in the hierarchy tree. Further inquiries about the membership
in the sets of feasible labels are then resolved within these
precomputed~sets.
 

%
%

%
%
%
%

\section{The Tool and Experiments}

The new CoIn input language has a~similar syntax to the standard DVE input
language of DiVinE, with the following differences. A~primitive automaton
(similar to DVE process)
consists of states and transitions and a~designed initial state. 
The transitions are labelled with triples.

\begin{center}
\begin{tabular}{|l|l|}
\hline
\begin{minipage}[t]{17em}
\footnotesize
\begin{verbatim}
automaton A (1) {
        state q0, q1, q2;
        init  q0;
        trans
                q0 -> q1 (1, a, -),
                q1 -> q2 (1, b, 1),
                q2 -> q0 (-, c, 1);
}
\end{verbatim}
\end{minipage}&
\begin{minipage}[t]{17em}
\footnotesize
\begin{verbatim}
automaton B (2) {
        state p0;
        init  p0;
        trans
                p0 -> p0 (-, a, 2),
                p0 -> p0 (2, c, -);
}
\end{verbatim}
\end{minipage}
\\\hline
\end{tabular}
\end{center}
\noindent A~composite automaton consists of a~set of automata and a~description
of the set of feasible labels. The description can be of two kinds --
\texttt{restrictL} denotes labels that are disallowed (all other labels
   are implicitly allowed), \texttt{onlyL} denotes labels that are allowed
(all other labels are implicitly disallowed).
The system automaton (the root of the hierarchy) is then denoted
using the \texttt{system} declaration.
\begin{center}
\begin{tabular}{|l|}
\hline
\begin{minipage}[t]{18em}
\footnotesize
\begin{verbatim}
composite C {
        A, B;
        restrictL (1, a, -), (-, c, 1);
}
system C;
\end{verbatim}
\end{minipage}\\\hline
\end{tabular}
\end{center}
\noindent
Augmenting the model with the never claim automaton corresponding to
a~given CI-LTL formula is done using an~external tool \texttt{coin-prop},
available at~\cite{coin-homepage}. The resulting file then may be verified
with CoInDiVinE.

The CoInDiVinE extension is a~part of the official
DiVinE distribution since version 2.5. 
The CoIn input language supports all standard algorithms
of DiVinE. It supports parallel and distributed
computation as well as the partial order reduction technique.
To perform the partial order reduction, 
we combine the topological sort proviso implemented in DiVinE~\cite{BBR10}
with our own heuristics for conditions C0--C2 tailored specifically for
the CI automata setting.

\begin{table}
\caption{State space generation on 1, 2, 4, 8, 16, 32 cores}
\label{table1}
\begin{center}
\small
\begin{tabular}{|c||r|r|r|r|r|r||r|r|r|r|r|r|}
\hline
\multirow{3}{*}{Model} &
\multicolumn{12}{c|}{\rule{0pt}{1em}time (s)}
\\\cline{2-13}
	& \multicolumn{6}{c||}{\rule{0pt}{1em}recursive}  
	& \multicolumn{6}{c|}{\rule{0pt}{1em}LCA} \\\cline{2-13}
  &\multicolumn{1}{c|}{\rule{0pt}{1em} 1 }
  &\multicolumn{1}{c|}{ 2 }
  &\multicolumn{1}{c|}{ 4 }
  &\multicolumn{1}{c|}{ 8 }
  &\multicolumn{1}{c|}{ 16 }
  &\multicolumn{1}{c||}{ 32 }
  &\multicolumn{1}{c|}{ 1 }
  &\multicolumn{1}{c|}{ 2 }
  &\multicolumn{1}{c|}{ 4 }
  &\multicolumn{1}{c|}{ 8 }
  &\multicolumn{1}{c|}{ 16 }
  &\multicolumn{1}{c|}{ 32}\\\hline
\rule{0pt}{1em}  SCM \hfill 2 & 2016 & 1071 & 555 & 284 & 192 & 134 & 1055 & 583 & 314 & 162 & 87 & 48\\\hline
\rule{0pt}{1em}  SCM \hfill 5 & 4193 & 2290 & 1152 & 601 & 346 & 284 & 2294 & 1156 & 675 & 358 & 192 & 106\\\hline
\rule{0pt}{1em}  SCR \hfill 2 & 276 & 144 & 73 & 38 & 23 & 18 & 144 & 79 & 42 & 21 & 11 & 6\\\hline
\rule{0pt}{1em}  SCR \hfill 5 & 591 & 308 & 159 & 81 & 49 & 42 & 315 & 178 & 97 & 50 & 26 & 15\\\hline
\rule{0pt}{1em}  TSC \hfill 2 & 125 & 65 & 33 & 17 & 11 & 9 & 62 & 35 & 18 & 9 & 5 & 3\\\hline
\rule{0pt}{1em}  TSC \hfill 5 & 264 & 140 & 72 & 37 & 22 & 18 & 140 & 80 & 42 & 22 & 12 & 7\\\hline
\end{tabular}

\bigskip

\begin{tabular}{|c||r|r|r|r|r|r||r|r|r|r|r|r|}
\hline
\multirow{3}{*}{Model} &
\multicolumn{12}{c|}{\rule{0pt}{1em}total memory (GB)}
\\\cline{2-13}
	& \multicolumn{6}{c||}{\rule{0pt}{1em}recursive}  
	& \multicolumn{6}{c|}{\rule{0pt}{1em}LCA} \\\cline{2-13}
  &\multicolumn{1}{c|}{\rule{0pt}{1em} 1 }
  &\multicolumn{1}{c|}{ 2 }
  &\multicolumn{1}{c|}{ 4 }
  &\multicolumn{1}{c|}{ 8 }
  &\multicolumn{1}{c|}{ 16 }
  &\multicolumn{1}{c||}{ 32 }
  &\multicolumn{1}{c|}{ 1 }
  &\multicolumn{1}{c|}{ 2 }
  &\multicolumn{1}{c|}{ 4 }
  &\multicolumn{1}{c|}{ 8 }
  &\multicolumn{1}{c|}{ 16 }
  &\multicolumn{1}{c|}{ 32}\\\hline
\rule{0pt}{1em}SCM \hfill 2 & 3.68 & 4.91 & 3.75 & 3.83 & 5.24 & 7.00 & 3.69 & 4.92 & 3.80 & 3.93 & 4.35 & 4.98\\\hline
\rule{0pt}{1em}SCM \hfill 5 & 7.32 & 11.07 & 7.46 & 7.61 & 9.26 & 13.93 & 7.33 & 11.03 & 7.53 & 7.72 & 8.65 & 9.61\\\hline
\rule{0pt}{1em}SCR \hfill 2 & 0.51 & 0.63 & 0.57 & 0.65 & 0.90 & 1.43 & 0.52 & 0.66 & 0.60 & 0.68 & 0.91 & 1.38\\\hline
\rule{0pt}{1em}SCR \hfill 5 & 0.99 & 1.50 & 1.07 & 1.16 & 1.58 & 2.55 & 1.00 & 1.51 & 1.11 & 1.21 & 1.47 & 2.00\\\hline
\rule{0pt}{1em}TSC \hfill 2 & 0.26 & 0.27 & 0.31 & 0.37 & 0.53 & 0.83 & 0.27 & 0.29 & 0.33 & 0.42 & 0.61 & 0.97\\\hline
\rule{0pt}{1em}TSC \hfill 5 & 0.47 & 0.50 & 0.54 & 0.62 & 0.79 & 1.16 & 0.48 & 0.51 & 0.57 & 0.65 & 0.86 & 1.27\\\hline
\end{tabular}
\end{center}
\end{table}

To evaluate the CoInDiVinE tool, we performed some experiments.
The model we used was the case study of~\cite{CoCoME}, complemented
with various usage scenarios, modelled as a~component automaton
describing the user. More details about the models used can be found
in~\cite{milan-diplomka}.
The experiments were done on a~$8\times8$-core Intel Xeon CPU X7560, 2.27GHz
machine. The results concerning time (in seconds) and maximum memory
usage (in GB) are presented in Table~\ref{table1}.

The experiments show two points. First, the LCA algorithm is significantly 
faster than the recursive algorithm. Second, both algorithms have good
scaling properties. This is mainly due to the fact that the
successor generating function is complex and takes most of the time.
Interestingly, the LCA algorithm scales better. This is
possibly due to the fact that the recursive algorithm allocates 
memory in each of its recursive invocations.
Moreover, the
memory overhead needed to store the precomputed information is negligible.
In cases with more threads and larger models, 
the maximum memory needed for the LCA algorithm is even smaller as the 
precomputed information is only stored in shared memory once.

As our tool supports partial order reduction, we also present 
experimental evaluation of the efficiency of this method in 
Table~\ref{table:por}. 
The experiments confirm our intuition that this technique is very efficient
in the context of hierarchical component-based systems
due to the high degree of interleaving of the components.
The reduction ratio is exactly the same when using the LCA or the recursive
algorithm and is also independent on the number of threads.

\begin{table}
\caption{Full state space versus reduced state space using 
  partial order reduction}
\label{table:por}
\begin{center}
\small
\begin{tabular}{|c|r|r|r|r|r@{ : }l|}
\hline
\multirow{2}{*}{Model} 	& \multicolumn{2}{c|}{Full} &  \multicolumn{2}{c|}{With p.o.r.} & \multicolumn{2}{c|}{reduction}\\\cline{2-5}
	& states & transitions & states & transitions & \multicolumn{2}{c|}{ratio}\\\hline
SCM \hfill 2&22\,745\,391&116\,949\,899&1\,808\,873&3\,334\,872 & 13&1\\\hline
SCM \hfill 5&45\,490\,782&275\,652\,425&3\,685\,976&8\,930\,215 & 12&1\\\hline
SCR \hfill 2&2\,994\,016&17\,015\,460&27\,437&58\,959  & 109&1\\\hline
SCR \hfill 5&5\,988\,032&44\,702\,380&55\,774&161\,166 & 107&1\\\hline
TSC \hfill 2&1\,356\,277&8\,222\,091&5\,495&8\,402    & \ \ 247&1 \\\hline
TSC \hfill 5&2\,712\,553&21\,746\,895&11\,253&23\,866 & 241&1 \\\hline
\end{tabular}
\end{center}
\end{table}

\bibliographystyle{eptcs}
\bibliography{biblio}
\end{document}